\documentclass[]{aastex63}
\usepackage{CJK}
\received{July 22, 2019}
\revised{October 31, 2019}
\accepted{November 6, 2019}
\submitjournal{ApJ}

\shorttitle{Evaluating the classification of Fermi BCUs}
\shortauthors{Kang et al.}
\begin{document}
 \begin{CJK*}{UTF8}{gbsn}
 
%\title{Evaluating the optical classification of Fermi BCUs using machine learning II: based on the Fermi Large Area Telescope Fourth Source Catalog}
%{(a preliminary version)} \footnote{Released on June, 10th, 2019}

\title{Evaluating the classification of Fermi BCUs from the 4FGL Catalog \\ Using Machine Learning}

\correspondingauthor{Shi-Ju Kang}\email{kangshiju@alumni.hust.edu.cn}

\author[0000-0002-9071-5469]{Shi-Ju Kang (康世举)}
\affil{School of Physics and Electrical Engineering, Liupanshui Normal University, Liupanshui, Guizhou, 553004, People's Republic of China}

\author{Enze Li}
\author{Wujing Ou}
\affiliation{School of Physics and Electrical Engineering, Liupanshui Normal University, Liupanshui, Guizhou, 553004, People's Republic of China}

\author{Kerui Zhu}
\affiliation{Department of Physics, Yunnan Normal University, Kunming, Yunnan, 650092, People's Republic of China}

\author{Jun-Hui Fan}
\affiliation{Center for Astrophysics, Guangzhou University, Guangzhou 510006, People's Republic of China}

\author[0000-0003-4773-4987]{Qingwen Wu}
\affiliation{School of Physics, Huazhong University of Science and Technology, Wuhan 430074, People's Republic of China}

\author{Yue Yin}
\affiliation{School of Physics and Electrical Engineering, Liupanshui Normal University, Liupanshui, Guizhou, 553004, People's Republic of China}

\collaboration{7}{}




\begin{abstract}
The recently published fourth Fermi Large Area Telescope source catalog (4FGL)\footnote{\url{https://arxiv.org/abs/1902.10045/}}\footnote{\url{https://fermi.gsfc.nasa.gov/ssc/data/access/lat/8yr_catalog/}}, reports 5065 gamma-ray sources in terms of direct observational gamma-ray properties. Among the sources, the largest population is the Active Galactic Nuclei (AGN), which consists of {3137} blazars, 42 radio galaxies, and {28} other AGNs. The blazar sample comprises {694}  flat-spectrum radio quasars (FSRQs), {1131}  BL Lac- type objects (BL Lacs), and {1312}  blazar candidates of an unknown type (BCUs). The classification of blazars is difficult using optical spectroscopy given the limited knowledge with respect to their intrinsic properties, and the limited availability of astronomical observations. To overcome these challenges, machine learning algorithms are being investigated as alternative approaches. Using the 4FGL catalog, a sample of {3137}  Fermi blazars with 23 parameters is systematically selected. Three established supervised machine learning algorithms (random forests (RFs), support vector machines (SVMs), artificial neural networks (ANNs)) are employed to general predictive models to classify the BCUs. We analyze the results for all of the different combinations of parameters. Interestingly, a previously reported trend the use of more parameters leading to higher accuracy is not found. Considering the least number of parameters used, combinations of {eight, 12 or 10} parameters in the {SVM, ANN, or RF} generated models achieve the highest accuracy (Accuracy $\simeq$ {91.8\%, or $\simeq$ 92.9\%}). Using the combined classification results from the optimal combinations of parameters, {724} BL Lac type candidates and  {332} FSRQ type candidates are  {predicted}; however, {256} {remain without a clear prediction}. 
\end{abstract}


%Galactic and extragalactic astronomy --- Galaxies --- Active galaxies --- Active galactic nuclei --- Blazars


\section{Introduction} \label{sec:Introduction}

{Some of the most} luminous sources in the extragalactic $\gamma$-ray sky are blazars, which are a sub-class of {active galactic nuclei} (AGN). Their multi-wavelength spectral energy distributions (SEDs) often exhibit a bimodal shape spanning the ${\rm log \nu-log \nu F_{\nu}}$ space and cover the entire electromagnetic spectrum (e.g., from radio to $\gamma$-ray bands). Their spectral energy is dominated by non-thermal emissions, and the origins of this energy are thought to be a relativistic jet at a small viewing angle with respect to the line of sight \citep{1995PASP..107..803U}. The lower energy part of the distribution is attributed to the synchrotron emission produced by the non-thermal electrons in the jet; this peaks within the millimeter to soft X-ray waveband. The higher energy part of the distribution is attributed to inverse Compton (IC) scattering; this peaks within the MeV to GeV range. Blazars are further classified into two categories, according to absence or presence of weak emission lines in their optical spectra \citep{1995PASP..107..803U}. This classification is based on the equivalent width (EW) of the emission line. BL Lacertae objects (BL Lacs) have weak or no emission lines; (e.g., ${\rm EW < 5 \AA}$ in the rest frame). Flat spectrum radio quasars (FSRQs) have  stronger emission lines (e.g., ${\rm EW\geq5 \AA}$). Blazars are also classified based on the peak frequency ($\nu^{\rm S}_{\rm p}$) of the lower energy bump in the SED, which is known as the synchrotron component. This classification approach identifies the sources as low (LSP, e.g., $\nu^{\rm S}_{\rm p}<10^{14}$ Hz), intermediate (ISP, e.g., $10^{14}~\rm Hz<\nu^{\rm S}_{\rm p}<10^{15}$ Hz), or high-synchrotron-peaked (HSP, e.g., $\nu^{\rm S}_{\rm p}>10^{15}$ Hz) blazars (e.g., \citealt{2010ApJ...716...30A,2016ApJS..226...20F}).

Recently, a new version of the fourth Fermi Large Area Telescope source catalog (4FGL) has been released  \citep{2019arXiv190210045T}. This version expands upon the first eight years of science data from the Fermi Gamma-ray Space Telescope mission in the energy range from 50 MeV to 1 TeV \citep{2019arXiv190210045T,2019arXiv190510771T}. The new 4FGL catalog includes 5065 sources above 4$\sigma$ significance, where most of these sources are AGNs. The largest source population of AGNs includes {3137} blazars (e.g., 98\%), 42 radio galaxies, and {28} other AGNs; More than 2860 AGNs are located at high Galactic latitudes ($|b|>10\arcdeg$). The {blazars}  in the 4FGL catalog include {694}  flat-spectrum radio quasars (FSRQs),  {1131} BL Lac- type objects (BL Lacs), and  {1312} blazar candidates of unknown type (BCUs).

Classified as FSRQs and BL Lacs in 4FGL catalog are the sources that their optical classifications have been well evaluated from the literature and/or optical spectrum
in the 4FGL catalog (see \citealt{2019arXiv190210045T,2019arXiv190510771T}). 
BCUs are the sources that {are classified} {as BZU} (blazars of uncertain type) object in the BZCAT catalog, and/or as a source displays a flat radio spectrum,  
a typical two-humped, blazar-like SED in one or more of the WISE, AT20G, VCS, CRATES, PMN-CA, CRATES-Gaps, or CLASS catalogs 
or in radio and X-ray catalogs (see \citealt{2019arXiv190210045T,2019arXiv190510771T} for the details and references therein).

{The Fermi catalog (each version, such as  First, Second, Third  and Fourth Fermi Large Area Telescope source catalogs, denoted as 1FGL, 2FGL, 3FGL or 4FGL) is by far the largest gamma ray source group (when released);}
it provides a unique and excellent opportunity for investigating the physics of the $\gamma$-ray emissions of blazars 
(e.g., \citealt{2012ApJ...753...45S,2014MNRAS.441.3375X,2015MNRAS.454..115S,2015MNRAS.451.2750X,2015MNRAS.450.3568X,2016RAA....16...13C,2016ApJS..226...20F,2016RAA....16..173F,2016Galax...4...36G,2016RAA....16..103L,2016ApJS..222...24X,2017RAA....17...66L, 2018ApJS..235...39C,2018RAA....18...56K,2018RAA....18..120L,2019ApJ...872..189K}.)

With respect to current techniques that have been used in the study of astronomy and astrophysics (e.g., \citealt{2019arXiv190407248B,2019arXiv190608349L,2019ApJ...881L...9F,2019ApJ...872..189K}), powerful data mining and machine learning algorithms have been widely adopted. 
Several literature reviews are available on this topic (\citealt{2010IJMPD..19.1049B}; \citealt{feigelson_babu_2012} and \citealt{2012amld.book.....W}).
The use of supervised machine learning (SML) algorithms, for example, has been extensively explored using the earlier versions of the Fermi catalogs (e.g., 1FGL, 2FGL, and 3FGL). 
For instance, using the 1FGL catalog, \cite{2012ApJ...753...83A} identified 221 AGN and 134 pulsar candidates from the 630 unassociated sources; this was accomplished using a random forest (RF) and a logical regression multivariate method. 
Using the 2FGL catalog, \cite{2012MNRAS.424L..64M} identified 216 AGN candidates from 269 unassociated sources (located at high galactic latitude $|b|>10\arcdeg$); this research adopted an RF method. 
\cite{2013MNRAS.428..220H} also used the 2FGL catalog; this research applied a support vector machine (SVM) and RF methods to identify BL Lac or FSRQ candidates from the 269 BCUs in 2FGL catalogue. 
In addition, \cite{2014ApJ...782...41D} applied an artificial neural network (ANN) and RF algorithms to identify “AGN” or “non-AGN” from 576 unassociated sources in the 2FGL catalog. 
Using the 3FGL catalog, \cite{2016MNRAS.462.3180C} identified 314 BL Lac candidates and 113 FSRQ candidates among the BCUs; this was accomplished using an ANN algorithm. 
\cite{2016ApJ...820....8S} also used the 3FGL catalog; this research identified 334 pulsar candidates and 559 AGN candidates from the unassociated sources using a RF and a logistic regression algorithm. 
\cite{2016Galax...4...14E} search for high-confidence Blazar Candidates based 3FGL, an infrared and an X-ray catalog using a RF algorithm.
In addition, \cite{2017A&A...602A..86L} firstly identified blazar candidates from the 3FGL unassociated sources; they subsequently classified BL Lacs or FSRQs from these candidates and the BCUs reported in 3FGL using multivariate classifications. 
Furthermore, \cite{2017MNRAS.470.1291S} used the 3FGL catalog and identified BL Lacs and FSRQs from 559 3FGL unassociated sources using an ANN algorithm.

{
Using} a 3LAC clean sample selected from the 3FGL catalog, \cite{2019ApJ...872..189K} (Paper I) identified BL Lacs and FSRQs using 4 different machine learning algorithms (Mclust Gaussian finite mixture models, Decision trees, RF, and SVM). {This research used eight parameters.}
In the work summarized above, only a subset of the parameters are selected for supervised learning based on certain selection criteria to evaluate the classification of the Fermi sources.
Taking a different approach, Xiao et al. (2019) has recently compiled an extensive collection of  parameters to utilize in the application of supervised learning. Even the astronomical coordinates of the Fermi source are considered in the initial stage of the supervised learning to evaluate the potential optical classification of BCUs. Based on the analysis of the research, in particular related to the selection of parameters, three open research questions are identified regarding the classification of blazars: RQ1 Do all possible parameters need to be considered in the application of SML algorithms?; RQ2 Does the use of more parameters improve the accuracy of SML algorithms?; RQ3 Does a single optimal combination of parameters exist for SML algorithms?

Three popular SML algorithms (RF, SVM, and ANN, e.g., see \citealt{2019arXiv190407248B} for the reviews) are utilized in this research to investigate the potential classification of BCUs. The 4FGL catalog is used, which provides the direct observational gamma-ray properties. Section 2 describes the method used to select the parameters and data sample from the catalog. In Section 3, the three SML techniques used in the research are briefly introduced. The classification results are reported in Section 4. Section 5 presents the discussion and conclusions of the research results.

\section{Sample} \label{sec:sample}

\begin{table*}
	\centering
	\caption{The result of two sample test for {1131} BL Lacs  and {694}  FSRQs}
	\label{tab:test}
	\begin{tabular}{clcccccccccccc} 
		\hline\hline
Paramaters &  Selected                  &\multicolumn{2}{c}{{KS test}}   & &\multicolumn{3}{c}{t-test} & &\multicolumn{2}{c}{Wilcox-test}    \\
\cline{3-4} \cline{6-8}	 \cline{10-11}			
{Label}&{Paramaters} & {$D$} & {$p_{1}$}&~& {$t$} &df& {$p_{2}$} &~& {$W$} & {$p_{3}$} &&{$Gini$} \\
		\hline
1	&	PL\_Index	&	0.76 	&	0	&	&	-43.92 	&	1495 	&	2.40E-271	&	&	51341 	&	7.18E-214	&	&	97.24 	\\
2	&	LP\_Index	&	0.70 	&	0	&	&	-37.06 	&	1697 	&	8.42E-221	&	&	69198 	&	2.81E-192	&	&	77.89 	\\
3	&	Pivot\_Energy	&	0.69 	&	0	&	&	26.57 	&	1504 	&	6.83E-128	&	&	720456 	&	6.76E-198	&	&	69.64 	\\
4	&	PLEC\_Flux\_Density	&	0.60 	&	0	&	&	-6.26 	&	701 	&	6.79E-10	&	&	108307 	&	4.93E-149	&	&	31.59 	\\
5	&	LP\_Flux\_Density	&	0.60 	&	0	&	&	-6.26 	&	701 	&	6.64E-10	&	&	108597 	&	9.84E-149	&	&	38.95 	\\
6	&	PL\_Flux\_Density	&	0.60 	&	0	&	&	-6.25 	&	702 	&	7.03E-10	&	&	110386 	&	6.86E-147	&	&	26.55 	\\
7	&	Frac2\_Variability	&	0.55 	&	0	&	&	-23.78 	&	1013 	&	1.15E-99	&	&	157701 	&	3.46E-109	&	&	39.45 	\\
8	&	nuFnu\_Band7	&	0.52 	&	0	&	&	7.84 	&	1166 	&	1.01E-14	&	&	636212 	&	3.38E-110	&	&	27.83 	\\
9	&	PLEC\_Index	&	0.50 	&	0	&	&	-19.85 	&	1555 	&	2.60E-78	&	&	158349 	&	8.47E-102	&	&	18.99 	\\
10	&	Flux\_Band2	&	0.50 	&	0	&	&	-6.61 	&	719 	&	7.51E-11	&	&	161629 	&	5.09E-99	&	&	16.53 	\\
11	&	Flux\_Band7	&	0.48 	&	0	&	&	7.79 	&	1195 	&	1.44E-14	&	&	622302 	&	3.40E-98	&	&	17.82 	\\
12	&	nuFnu\_Band2	&	0.48 	&	0	&	&	-6.54 	&	720 	&	1.18E-10	&	&	166273 	&	3.75E-95	&	&	16.58 	\\
13	&	Frac\_Variability	&	0.47 	&	0	&	&	-21.77 	&	1143 	&	3.72E-88	&	&	168016 	&	2.72E-94	&	&	19.07 	\\
14	&	PLEC\_Expfactor	&	0.47 	&	0	&	&	-13.53 	&	826 	&	8.08E-38	&	&	186454 	&	2.96E-79	&	&	21.83 	\\
15	&	Variability2\_Index	&	0.44 	&	0	&	&	-3.96 	&	701 	&	8.30E-05	&	&	182992 	&	7.07E-82	&	&	13.22 	\\
16	&	Variability\_Index	&	0.42 	&	0	&	&	-4.56 	&	708 	&	6.16E-06	&	&	187942 	&	3.85E-78	&	&	14.01 	\\
17	&	nuFnu\_Band6	&	0.41 	&	0	&	&	6.13 	&	1716 	&	1.09E-09	&	&	582328 	&	1.31E-67	&	&	17.58 	\\
18	&	Flux\_Band3	&	0.40 	&	0	&	&	-5.83 	&	743 	&	8.35E-09	&	&	193086 	&	2.36E-74	&	&	8.03 	\\
19	&	Npred	&	0.40 	&	0	&	&	-7.63 	&	840 	&	6.38E-14	&	&	199184 	&	5.50E-70	&	&	7.58 	\\
20	&	Flux\_Band6	&	0.39 	&	0	&	&	5.74 	&	1767 	&	1.11E-08	&	&	575155 	&	9.83E-63	&	&	14.38 	\\
21	&	nuFnu\_Band3	&	0.38 	&	0	&	&	-5.68 	&	747 	&	1.95E-08	&	&	201640 	&	2.89E-68	&	&	6.97 	\\
22	&	Flux\_Band1	&	0.32 	&	0	&	&	-6.34 	&	716 	&	4.00E-10	&	&	239629 	&	1.95E-44	&	&	7.31 	\\
23	&	nuFnu\_Band1	&	0.32 	&	0	&	&	-6.33 	&	716 	&	4.41E-10	&	&	240708 	&	7.78E-44	&	&	8.14 	\\
\hline
24	&	LP\_beta	&	0.21 	&	1.11E-16	&	&	-3.98 	&	1466 	&	7.14E-05	&	&	309584 	&	3.38E-14	&	&	8.32 	\\
25	&	LP\_SigCurv	&	0.20 	&	9.99E-16	&	&	-9.25 	&	912 	&	1.49E-19	&	&	291625 	&	2.80E-20	&	&	8.18 	\\
26	&	Energy\_Flux100	&	0.19 	&	3.10E-13	&	&	-3.78 	&	858 	&	1.67E-04	&	&	291490 	&	2.50E-20	&	&	6.22 	\\
27	&	PLEC\_SigCurv	&	0.17 	&	2.15E-11	&	&	-8.28 	&	919 	&	4.44E-16	&	&	308207 	&	1.27E-14	&	&	7.75 	\\
28	&	Flux\_Band4	&	0.17 	&	2.22E-11	&	&	-3.99 	&	812 	&	7.08E-05	&	&	302615 	&	2.02E-16	&	&	5.94 	\\
29	&	nuFnu\_Band4	&	0.15 	&	2.35E-09	&	&	-3.78 	&	824 	&	1.65E-04	&	&	312294 	&	2.22E-13	&	&	5.84 	\\
30	&	nuFnu\_Band5	&	0.15 	&	5.52E-09	&	&	0.08 	&	1237 	&	9.40E-01	&	&	456681 	&	4.19E-09	&	&	7.97 	\\
31	&	Flux\_Band5	&	0.14 	&	6.15E-08	&	&	-0.41 	&	1159 	&	6.82E-01	&	&	445158 	&	1.42E-06	&	&	7.17 	\\
32	&	Flux1000	&	0.11 	&	3.40E-05	&	&	-3.03 	&	881 	&	2.55E-03	&	&	347260 	&	3.54E-05	&	&	6.35 	\\
33	&	Signif\_Avg	&	0.11 	&	6.06E-05	&	&	-3.79 	&	1110 	&	1.57E-04	&	&	362770 	&	6.60E-03	&	&	7.61 	\\
34	&	PLEC\_Exp\_Index	&	0.00 	&	1.000 	&	&	1.00 	&	693 	&	0.32 	&	&	393023 	&	0.20 	&	&	0.01 	\\
\hline
	\end{tabular}
	\\
\tablecomments{Column 1 presents the parameter labels in the sample. 
Column 2 lists the selected parameters.
The  two-sample Kolmogorov−Smirnov test results for the test statistic ($D$) and the p-value($p_1$) are presented in Columns 3 and 4, respectively. 
The Welch Two Sample t-test results for  the t-statistic($t$), the degrees of freedom for the t-statistic (df), and the p-value($p_2$) are presented in Columns 5, 6, and 7, respectively. 
The Wilcoxon rank sum test with continuity correction test statistic ($W$) and p-value ($p_3$) are presented in Columns 8 and 9, respectively. 
The Gini coefficient ($Gini$), an indicator of variable importance in RFs are presented in Column 10.}
\end{table*}

In the updated table released on {30 September 2019} (4FGL FITS file { ``gll\_psc\_v20.fit"\footnote{\url{https://fermi.gsfc.nasa.gov/ssc/data/access/lat/8yr_catalog/gll_psc_v20.fit}}}) of the 4FGL catalog, {84} variables are reported using {333} columns (\citealt{2019arXiv190210045T}, {Table 12}). 
 {Among the {84}  variables, some variables contain multiple columns.} {The data} includes the parameters  ``Flux\_Band" (seven columns are used to present the integral photon flux in each of the seven spectral bands {that are marked as Flux\_Band1, Flux\_Band2 ... respectively, see Table \ref{tab:test}}); “Unc\_Flux\_Band” (14 columns are used to present the 1$\sigma$  lower and upper error for each  ``Flux\_Band"); “nuFnu{\_Band}” (seven columns are used to present the SED for the spectral bands, {marked as nuFnu{\_Band}1, nuFnu{\_Band}2 ... respectively, see Table \ref{tab:test}}); “Sqrt\_TS\_Band (seven columns are used to represent the square root of the test statistic for the spectral bands). Historical parameters include the ``Flux\_History", ``Unc\_Flux\_History" and ``Sqrt\_TS\_History", which use 8, 16, and 8 columns, respectively to present the annual integral photon flux from 100 MeV to 100 GeV, 1$\sigma$ lower and upper error on the integral photon flux, and the square root of the test statistic; 
{and the ``Flux2\_History", ``Unc\_Flux2\_History" and ``Sqrt\_TS2\_History", which use 48, 96, and 48 columns, respectively to present the each two-moth integral photon flux from 100 MeV to 100 GeV, 1$\sigma$ lower and upper error on the integral photon flux, and the square root of the test statistic}.

There are three main steps that are used to create the data sample. The first step is to identify the subset of parameters and their associated data. To achieve this, the coordinate columns, error columns, string columns, and most data missing columns (e.g.,“Sqrt TS Band”) are removed. This results in selecting {34} candidate parameters (see Table \ref{tab:test}) from the 4FGL table. In order to simplify the calculation, some parameters are pre-selected for the SML algorithms. To accomplish this, three two sample tests are used to calculate the independence of the {34} parameters. These tests are the Kolmogorov-Smirnov test, Welch Two Sample t-test, and Wilcoxon rank sum test with continuity correction (e.g., \citealt{2018MNRAS.475.1708A,2019ApJ...872..189K}). The tests are applied to two subsamples of the data ({694}  FSRQs and {1131} BL Lacs); the results are summarized in Table \ref{tab:test}.
Considering p $>$ 0.05\footnote{where, p $>$ 0.05 indicates that the two populations should be the same distribution, which does not reject the null hypothesis},
one parameter (``PLEC\_Exp\_Index") is excluded; therefore,  {33} parameters are selected in this work.

The second step is to ensure the parameter selection is reasonable. This is accomplished by applying a RF algorithm (see Section  \ref{sec:method}) to the parameters’ data to compute the Gini coefficients {(see \citealt{randomForest}; \citealt{Breiman2003Random} for the details and references therein)}, which is an established method to determine the variables’ importance. These results are presented in Table \ref{tab:test}; they are consistent with those of the two sample tests. Based on the  selected  {33} parameters, a subset of the data is selected from the 4FGL catalog, which includes {3137} blazars ({1131} BL Lacs, {1312} BCUs, and {694}  FSRQs).
The catalog has a total of {1312} BCUs; these are listed in Table \ref{ML_result}.

The third step is to further reduce the number of parameters, to ensure the study  { can be completed}. For the selected {33} parameters, there are {8589934591} different combinations. 
{Assuming that we utilize the RF, SVM, and ANN to calculate each combination of parameters, the computer (with 4 cores) requires approximately 1/4 second to accomplish this.
In this scenario, computing all of the combinations costs approximately {24855} days, i.e., over {68} years.}
This is not possible for us to accomplish. In order to reduce the calculation time, we further sub-selected 23 parameters (8388607 different combinations, requiring approximately 24 days) to carry out our work. This sub-selection is also based on the two sample test results (see Table \ref{tab:test}), by considering D $>$ {0.30} in the Kolmogorov-Smirnov test, or $p_3 < 1.00E-{20}$  in the Wilcoxon rank sum test. A simple horizontal line is introduced in the table to distinguish the collection of 23 parameters utilized.

\section{supervised machine learning algorithms} \label{sec:method}

In this section, a brief introduction to the popular, well established SML algorithms that are adopted in this work is provided, including RF, SVM, and ANN (e.g., see \citealt{2019arXiv190407248B} for the reviews). These approaches share a common, high level approach to establish and assess the accuracy of predictive models. They divide a dataset into training, validation, and forecast samples. The training and validation samples contains predictive variables and known outcomes; the forecast sample only contain predictive variables.  SML algorithms use the training sample to generate a predictive model; the accuracy of the model is evaluated using the separate validation sample. An accurate model can subsequently be used to predict the outcomes of the forecast datasets, for which the outcomes are not known a priori (\citealt{feigelson_babu_2012,Kabacoff2015R}). 

Beyond the high level, common approach of dividing the dataset to establish and assess predictive models, the RF, SVM, and ANN algorithms have distinct characteristics. The original RF proposal (\citealt{Breiman2001}), which has evolved over time, transforms a training sample into a large collection of decisions trees (e.g., a forest). These trees are used to conduct an extensive voting scheme, which enhances the classification and the prediction accuracy of the model. The RF algorithm has numerous advantages, including accuracy, scalability, and the ability to address challenging datasets. In terms of accuracy, the RF approach has outperformed alternative approaches, for example, decision trees (\citealt{JMLRv15delgado2014a}). This approach has been applied to a very large astronomical dataset (\citealt{Breiman2003Random}). RF successfully builds predictive models for uneven datasets, for example, those with large amounts of missing data or a relatively limited ratio of observations in comparison to the number of variables. The RF approach also generates out-of-bag error rates, in addition to measures indicating the relative importance of the variables.

The original SVM proposal ({Vladimir N. Vapnik and Alexey Ya. Chervonenkis, 1963}) has evolved over the years, providing improvements in accuracy and performance. In the recent versions of the algorithm (\citealt{Vapnik1995,Vapnik2000}), data points in a training sample are efficiently mapped into a hyperplane, or a set of hyperplanes. The training sample is often defined in a finite-dimensional problem space, whereas the hyperplanes are defined in a high- or infinite-dimensional space to make their separation easier. The hyperplanes are defined as an orthogonal, and therefore minimal, set of vectors. More specifically, the dot product of two training data points with a vector in that space is a constant value. The hyperplanes achieve the largest functional margin to improve the accuracy of the model, as this reflects greater distances that separate the classes. The SVM algorithms are popular; they have been used to generate accurate prediction models in a wide variety of domains.

The original ANN proposal (\citealt{ANNcite}) has also been improved over time; they are a family of algorithms based on structures that are inspired by biological neural networks, such as the human brain. These structures consist of multiple layers, including an input layer, one or more hidden layers, and an output layer. Each layer is designed to recognize a specific element in the data, then propagate the result to the next layer. In combination, the layers can learn to recognize complex features in the data. A key advantage of ANNs is that the network can actually learn from observing data sets and become more accurate. In this way, ANN is used as a random function approximation tool. However, they require large training sets and have high computational demands.

Numerous software packages are available for the RF, SVM, and ANN algorithms. The following have been selected for this work. 
For the RF algorithm, the randomForest package (\citealt{randomForest}) in R\footnote{\url{https://www.R-project.org/}} {(R version 3.6.1, \citealt{RLanguage})} is used to fit the random forests. 
For the SVM algorithm, the e1071 package (\citealt{SVMe1071}) is used to build the predictive model in the base R installation. 
For the ANN algorithm, the nnet packages (\citealt{ANNcite}) in R are used to create the predictive network.
In addition, the accuracy of the models is calculated using the $classAgreement()$ function in the e1071 package.

\section{Results} \label{sec:result}

%%% Table 2
\begin{deluxetable*}{ccccccccccccccccc}
\tablenum{2}
\tablecaption{The test accuracy, predict results , and parameters for the optimal combinations}
\tablewidth{0pt}
\tablehead{
\colhead{classifier} & \colhead{Number} & \colhead{bll} & \colhead{fsrq} & \colhead{Accuracy} & 
\colhead{par1} & \colhead{par2} &\colhead{par3} & \colhead{par4} &\colhead{par5} & \colhead{par6} &
\colhead{par7} & \colhead{par8} &\colhead{par9} & \colhead{par10} &\colhead{par11} &\colhead{par12} 
}
\decimalcolnumbers
\startdata
%-------------------------------------------------------------------------------------------------------------------------------------------------------
{\bf ANN} &	12	&	868	&	444	&	0.918 	&	1	&	4	&	5	&	6	&	7	&	8	&	9	&	17	&	19	&	21	&	22	&	23	\\
{\bf RF}    &	10	&	835	&	477	&	0.929 	&	1	&	3	&	7	&	8	&	10	&	11	&	13	&	14	&	17	&	22	&		&		\\
{\bf SVM} &	8	&	858	&	454	&	0.918 	&	3	&	8	&	10	&	11	&	12	&	15	&	16	&	17	&		&		&		&		\\
SVM 	&	8	&	859	&	453	&	0.918 	&	3	&	8	&	10	&	11	&	12	&	15	&	16	&	20	&		&		&		&		\\
\bf\emph{SVM}&8	&	817	&	495	&	0.918 	&	3	&	8	&	10	&	14	&	16	&	19	&	20	&	22	&		&		&		&		\\
SVM 	&	8	&	816	&	496	&	0.918 	&	3	&	8	&	10	&	14	&	16	&	19	&	20	&	23	&		&		&		&		\\
SVM 	&	8	&	812	&	500	&	0.918 	&	3	&	8	&	11	&	12	&	14	&	16	&	17	&	23	&		&		&		&		\\
SVM 	&	8	&	816	&	496	&	0.918 	&	3	&	8	&	12	&	14	&	16	&	19	&	20	&	22	&		&		&		&		\\
SVM 	&	8	&	817	&	495	&	0.918 	&	3	&	8	&	12	&	14	&	16	&	19	&	20	&	23	&		&		&		&		\\
SVM 	&	8	&	815	&	497	&	0.918 	&	3	&	10	&	11	&	14	&	16	&	17	&	19	&	22	&		&		&		&		\\
SVM 	&	8	&	817	&	495	&	0.918 	&	3	&	10	&	11	&	14	&	16	&	17	&	19	&	23	&		&		&		&		\\
SVM 	&	8	&	817	&	495	&	0.918 	&	3	&	10	&	11	&	14	&	16	&	19	&	20	&	22	&		&		&		&		\\
SVM 	&	8	&	819	&	493	&	0.918 	&	3	&	10	&	11	&	14	&	16	&	19	&	20	&	23	&		&		&		&		\\
SVM 	&	8	&	816	&	496	&	0.918 	&	3	&	11	&	12	&	14	&	16	&	17	&	19	&	22	&		&		&		&		\\
SVM 	&	8	&	817	&	495	&	0.918 	&	3	&	11	&	12	&	14	&	16	&	17	&	19	&	23	&		&		&		&		\\
SVM 	&	8	&	818	&	494	&	0.918 	&	3	&	11	&	12	&	14	&	16	&	19	&	20	&	22	&		&		&		&		\\
SVM 	&	8	&	819	&	493	&	0.918 	&	3	&	11	&	12	&	14	&	16	&	19	&	20	&	23	&		&		&		&		\\%-------------------------------------------------------------------------------------------------------------------------------------------------------
\enddata
\tablecomments{The classifiers are presented in Column 1 and the number of parameters for the optimal combination (with least parameters) are presented in Column 2. The number of BL Lacs and FSRQs predicted by a supervised classifier (using SML techniques) for the BCUs (predicted dataset) are presented in Columns 3 and 4. The highest accuracies of each classifier are presented in Column 5. The labels of the parameters are presented in Columns 6-17; these correspond to the labels in Table \ref{tab:test}, Column 1. Here, the least parameters for the optimal combination is shown for guidance regarding its form and content. The complete results are available elsewhere in a machine readable format (see Table2\_combination.xlsx).}
\label{tab_resultA}
\end{deluxetable*}

\begin{figure*}[bpt]
\centering
        \includegraphics[height=11cm,width=18cm]{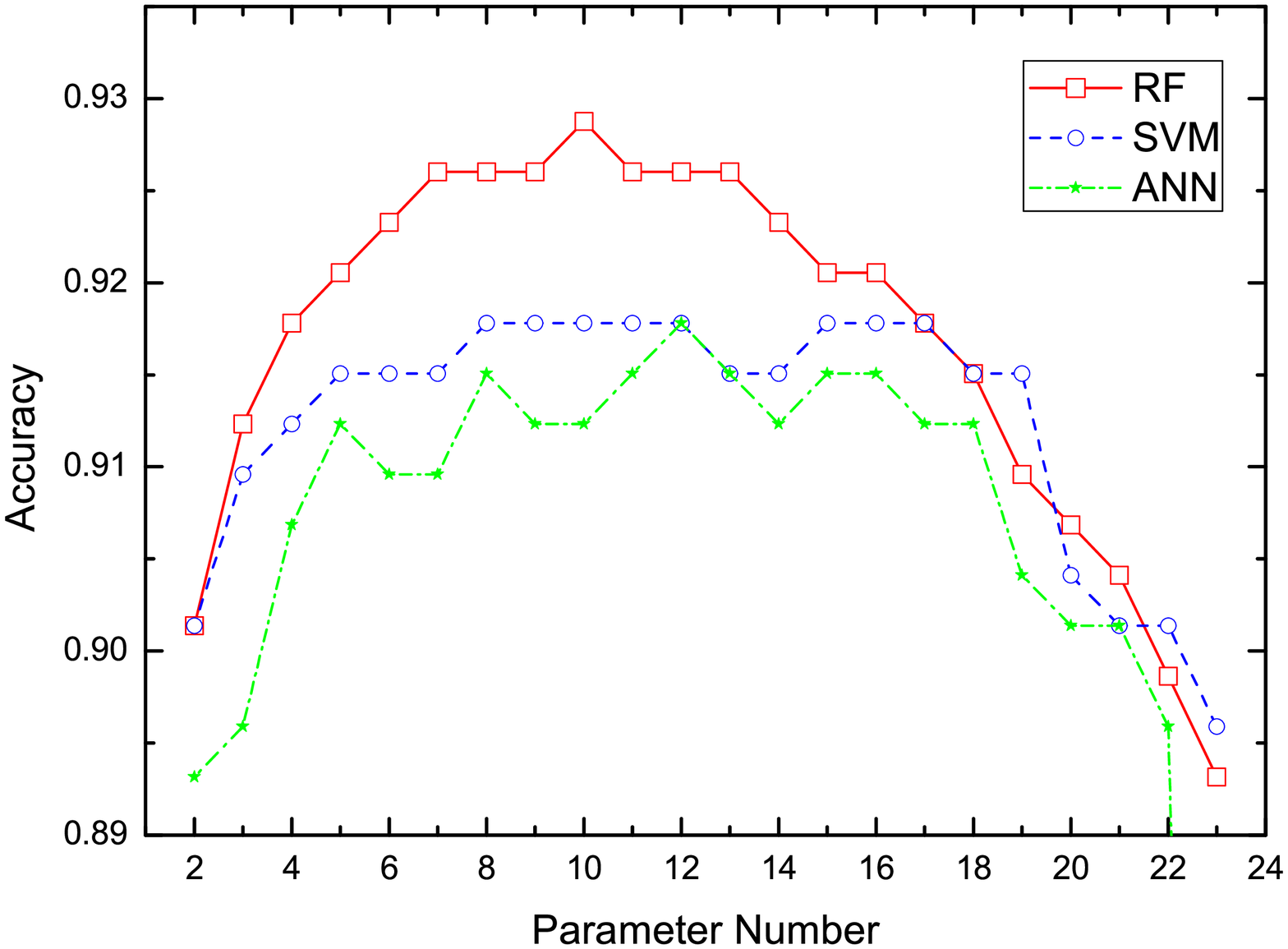}
\caption{(Color online) Highest accuracy for the different number of combinations of parameters in each SML algorithm: Random Forests (red solid line + empty square), Support Vector Machines (blue dashed line + empty circle), and Artificial Neural Networks (green dotted-dashed line + star).}
 \label{sub:figa}
\end{figure*}

\begin{deluxetable*}{llcccccccccc}
\tiny
\tablenum{3}
\tablecaption{The classification of $Fermi$ BCUs \label{ML_result}}
\tablewidth{0pt}
\tablehead{
\colhead{4FGL Name}           & \colhead{Counterpart name}     &\colhead{${\rm C_{4FGL}}$}  & 
\colhead{${\rm C_{RF}}$}       & \colhead{$\rm P_{Bi,RF}$}     & \colhead{$\rm P_{Fi,RF}$}
& \colhead{${\rm C_{SVM}}$} & \colhead{$\rm P_{Bi,SVM}$}     & \colhead{$\rm P_{Fi,SVM}$}
& \colhead{${\rm C_{ANN}}$} & \colhead{$\rm P_{Bi,ANN}$}     & \colhead{$\rm P_{Fi,ANN}$}
}
\decimalcolnumbers
\startdata
%Source_Name	&	ASSOC1	&	CLASS1	&	RF_result_order	&	bll	&	fsrq	&	SVM_result_order	&	bll	&	fsrq	&	ANN_result_order	&	bll	&	fsrq	\\
4FGL J0001.2$+$4741 	&	B3 2358$+$474                     	&	bcu	&	bll	&	0.670 	&	0.330 	&	bll	&	0.908 	&	0.092 	&	bll	&	0.899 	&	0.101 	\\
4FGL J0001.6$-$4156 	&	1RXS J000135.5$-$415519       	&	bcu	&	bll	&	1.000 	&	0.000 	&	bll	&	0.993 	&	0.007 	&	bll	&	0.899 	&	0.101 	\\
4FGL J0002.1$-$6728 	&	SUMSS J000215$-$672653        	&	bcu	&	bll	&	0.972 	&	0.028 	&	bll	&	0.994 	&	0.006 	&	bll	&	0.899 	&	0.101 	\\
4FGL J0003.1$-$5248 	&	RBS 0006                              	&	bcu	&	bll	&	1.000 	&	0.000 	&	bll	&	0.994 	&	0.006 	&	bll	&	0.899 	&	0.101 	\\
4FGL J0003.3$-$1928 	&	PKS 0000$-$197                    	&	bcu	&	bll	&	0.588 	&	0.412 	&	fsrq	&	0.336 	&	0.664 	&	fsrq	&	0.112 	&	0.888 	\\
4FGL J0003.3$-$5905 	&	PMN J0003$-$5905                	&	bcu	&	bll	&	0.650 	&	0.350 	&	fsrq	&	0.372 	&	0.628 	&	bll	&	0.899 	&	0.101 	\\
4FGL J0004.0$+$0840 	&	SDSS J000359.23$+$084138.1    	&	bcu	&	bll	&	0.890 	&	0.110 	&	bll	&	0.854 	&	0.146 	&	bll	&	0.899 	&	0.101 	\\
4FGL J0006.4$+$0135 	&	NVSS J000626$+$013611         	&	bcu	&	bll	&	0.970 	&	0.030 	&	bll	&	0.920 	&	0.080 	&	bll	&	0.899 	&	0.101 	\\
4FGL J0007.7$+$4008 	&	NVSS J000741$+$400830         	&	bcu	&	bll	&	0.984 	&	0.016 	&	bll	&	0.779 	&	0.221 	&	bll	&	0.899 	&	0.101 	\\
4FGL J0008.0$-$3937 	&	PMN J0008$-$3945               	&	bcu	&	fsrq	&	0.290 	&	0.710 	&	fsrq	&	0.226 	&	0.774 	&	fsrq	&	0.112 	&	0.888 	\\
4FGL J0008.4$+$1455 	&	NVSS J000825$+$145635         	&	bcu	&	bll	&	0.898 	&	0.102 	&	bll	&	0.745 	&	0.255 	&	bll	&	0.899 	&	0.101 	\\
4FGL J0010.8$-$2154 	&	PKS 0008$-$222                   	&	bcu	&	fsrq	&	0.386 	&	0.614 	&	fsrq	&	0.286 	&	0.714 	&	bll	&	0.899 	&	0.101 	\\
4FGL J0011.8$-$3142 	&	SUMSS J001141$-$314220        	&	bcu	&	bll	&	0.904 	&	0.096 	&	bll	&	0.980 	&	0.020 	&	bll	&	0.899 	&	0.101 	\\
4FGL J0012.0$+$7043 	&	TXS 0008$+$704                    	&	bcu	&	fsrq	&	0.226 	&	0.774 	&	bll	&	0.581 	&	0.419 	&	fsrq	&	0.112 	&	0.888 	\\
…	&	…	&	…	&	…	&	…	&	…	&	…	&	…	&	…	&	…	&	…	&	…	\\\enddata
\tablecomments{The 4FGL names are presented in Column 1 and the counterpart names are presented in Column 2. The optical classes are presented in Column 3 (e.g., BCU as reported in \citealt{2019arXiv190210045T}). The classification of the $Fermi$ BCUs using the Random Forest (${\rm C_{RF}}$), Support Vector Machines (${\rm C_{SVM}}$), and Artificial neural networks (${\rm C_{ANN}}$) are presented in Columns 4, 7`, and 10, where “bll” indicates BL Lac and “fsrq” indicates FSRQ. 
{The probabilities $P_{Bi,X}$ and $P_{Fi,X}$ that a BCU i belongs to the BL Lac or FSRQ classes 
from the RF method are shown in Columns 5 and 6; from the SVM method are shown in Columns 8 and 9; and from the ANN method are shown in Columns 11 and 12; respectively.}
The complete results are available elsewhere in machine readable format Table \ref{ML_result} is published in its entirety in the machine-readable format (see Table3\_classification.xlsx). 
A portion is shown here for guidance regarding its form and content.}
\end{deluxetable*}

The selected dataset, consisting of {3137} Fermi sources (blazars) with {694}  FSRQs, {1131} BL Lacs, and {1312} BCUs, are divided into three samples: training, validation, and forecast. The FSRQs and the BL Lacs have known optical classifications. {Approximately 4/5} of these are randomly assigned to the training sample; the random seed value of 123 is used. The remaining data (e.g., {approximately 1/5}) are used as the validation dataset for different combinations of parameters in each of the SML algorithms (SVM, RF, and ANN). In summary the training dataset includes  {1460} blazars ({905} BL Lacs, {555} FSRQs), and the validation dataset contains  {365} blazars (\emph{SVM, RF, and ANN}). The forecast dataset consists of the {1312} BCUs. 
The training, validation, and forecast samples are used by the SVM, RF, and ANN algorithms.
The default settings for each of the three classification functions ($svm()$, $randomForest()$, and $nnet()$) are used to simplify the calculations. After the predictive models are generated and assessed; an effective predictive model is used to forecast whether a BCU belongs to the FRSQ or the BL Lac class based on its predictor variables. The main steps to accomplish this in the R platform are publicly available\footnote{\url{https://github.com/ksj7924/Kang2019ApJRcode}}.

The accuracies of the different parameter combinations in each of the SML algorithms (\emph{SVM, RF, and ANN}) are computed.
The highest prediction accuracies for different combinations of parameters in each of the SML algorithms are illustrated in Figure \ref{sub:figa}. In the figure, the RF is represented with a red solid line + empty square; SVM is represented with a blue dashed line + empty circle,  and ANN is represented with a green dotted-dashed line + star. As the number of parameters increases, the accuracy gradually reaches its maximum. 
Here, with {eight, 10, or 12} parameters combinations (see Table \ref{tab_resultA}), 
the accuracy of the SVM, RF, or ANN reaches its maximum, respectively. 
For the ANN algorithm, there is {one combination} with 12 parameters achieving a maximum accuracy (accuracy={0.918}). 
For the RF algorithm, there is one combination of 10 parameters achieving a maximum accuracy (accuracy={0.929}). 
For the SVM algorithm, there are {15} combinations with {8} parameters that achieve a maximum accuracy (accuracy={0.918})(see Table \ref{tab_resultA}). 
{These results reveal that there are different combinations with the same number of parameters achieving a maximum accuracy in SVM algorithm.}
When more parameters are applied, the accuracy begins to decline (e.g., {12, 10, or 17} parameters combination in ANN, RF, or SVM). 
These results are inconsistent with the conclusions of our previous work (Paper I) that showed more parameters and more accuracy.

According to the optimal combination of parameters ({eight, 10, or 12} parameters in {SVM, RF, or ANN}), an optimized prediction model is obtained from each of the SML algorithms (e.g., {SVM, RF, or ANN}). These optimized models are used to predict whether a BCU belongs to the BL Lacs or the FSRQs (see Table \ref{tab_resultA} and Table \ref{ML_result}). 
Using the model generated by the RF algorithm, {835} BL Lac style candidates and {477} FSRQ style candidates are obtained from an optimal combination of 10 parameters (see Table 2) with a maximum accuracy of {0.929}. 
Using the model generated by the ANN algorithm, there are {one combination with 12} parameters (see Table \ref{tab_resultA}) with a maximum accuracy of {0.918}), where, {868} BCU sources are diagnosed as BL Lac type candidates. The remaining {444} BCU sources are diagnosed as FSRQ type candidates. 
However, for the SVM SML algorithm, there are {15} combinations with {8} parameters (see Table \ref{tab_resultA}) with a maximum accuracy of {0.918}). 
These combinations result in the diagnosis of {858, 859, 817, 816, 812, 816, 817, 817, 815, 817, 819, 817, 816, 818, or 819} BL Lac style candidates, respectively (see Table 2). 
The remaining {454, 453, 495, 496, 500, 496, 495, 495, 497, 495, 493, 495, 496, 494, or 493} BCUs are diagnosed as FSRQ style candidates.

{A better predictions can be obtained by applying various algorithms simultaneously (see Paper I).}
Based on the optimal combinations of the least number of parameters (e.g., eight parameters  in SVM), the combined the results of RF and ANN and one of SVM (with boldface letter, see Table \ref{tab_resultA}) demonstrate {724} BL Lac type candidates and {332} FSRQ type candidates are consistently predicted. However, there are {256} uncertain type BCUs (unks), for which there are inconsistent results.

%2019.06.30
\section{Discussions and Conclusions} \label{sec:Disc}

 The potential classification of Fermi BCUs is investigated in this work using three established  SML algorithms (\emph{RF, SVM and ANN});  the optimal combination of parameters is sought. Based on the 4FGL catalog, {3137} Fermi blazars with 23 parameters are selected from the 4FGL fits table (see Section \ref{sec:sample}, Section \ref{sec:result}, Table \ref{tab:test}). Using the algorithms, optimized predictive models are generated and the classification of BCUs based on their directly observable gamma-ray properties is conducted. For the 23 parameters, all of the possible combinations of the parameters are considered. We find that the previously reported trend of using more parameters resulting in more accuracy is not consistently presented. Combinations of {eight (with least parameters), 12, or 10} parameters  in SVM, RF, or ANN algorithms, respectively, achieve the highest accuracies (accuracy $\simeq$ {91.8}\%, or $\simeq$ {92.9}\%). Based on the optimal combination of parameters, the combined the results (indicated with boldface, see Table \ref{tab_resultA}) of these three algorithms result in the diagnosis of {724} BL Lac type candidates and {332} FSRQ type candidates. However, there are {256} uncertain type BCUs forecast.

The different physical origins of $\gamma$-ray emissions in BL Lacs and FSRQs may account for their distinct gamma-ray properties 
(e.g., \citealt{2016RAA....16...54B,2016RAA....16..173F,2018SCPMA..61e9511Y,2019MNRAS.482L..80B}).
For BL Lacs, the $\gamma$-ray emission is generally believed to originate from a pure synchrotron self-Compton (SSC) process 
(e.g., \citealt{1997A&A...320...19M}; \citealt{2004ApJ...601..151K}; \citealt{2014ApJ...788..104Z}),
while for FSRQs, the $\gamma$-ray emission is generally believed to originate from the SSC + EC (external- Compton) radiation process 
(e.g., \citealt{1999ApJ...515..140S}; \citealt{2002ApJ...581..127B}; \citealt{2014MNRAS.439.2933Y}; \citealt{2011MNRAS.414.2674G}; \citealt{2011ApJ...735..108C}; \citealt{2017ApJS..228....1Z}).
This indicates the FSRQ jet has a complex external physical environment, which may lead to complex physical properties in Fermi energy bands 
(e.g., see \citealt{2019ApJ...872..189K} for similar discussions).
For instance, the photon spectral index of FSRQs is greater than that the index of BL Lacs 
(see \citealt{2010ApJ...715..429A,2011ApJ...743..171A,2015ApJ...810...14A}; \citealt{2019arXiv190210045T,2019arXiv190510771T})
which may be the result of the spectrum being superposed with other spectra components for the FSRQs 
(e.g., \citealt{2013ApJ...764..113Z,2014ApJS..215....5K}; \citealt{2016A&A...585A...8Z}; \citealt{2016MNRAS.461.1862K,2017ApJ...837...38K}).
Therefore, the Fermi band of FSRQs, located at the intersection of both the SSC component and the EC component, may  result in more complex observational features 
(e.g., a hard spectrum, \citealt{2009ApJ...700..597A,2010ApJ...715..429A}; \citealt{2011ApJ...743..171A,2015ApJ...810...14A}).
At a deeper level, intrinsically different physical origins between the FSRQs and BL Lacs need further investigation. For instance, research directions include the different accretion models 
 (e.g., see \citealt{2002ApJ...579..554W,2003ApJ...599..147C,2009MNRAS.396L.105G,2009ApJ...694L.107X,2014MNRAS.445...81S,2015AJ....150....8C,2018ApJS..235...39C,2018MNRAS.473.2639G} for more details and reference therein),
where FSRQs have a standard cold accretion disk and BL Lacs have an advection-dominated accretion flow (ADAF; Yuan \& Narayan 2014);  
the mass accretion rate on the central black hole (e.g., see \citealt{2019MNRAS.482L..80B});
and the mass accretion rate and the magnetic field strength 
(e.g., see \citealt{2019MNRAS.tmp.1011M}).

According to the results for the different combinations of parameters, although the prediction accuracies are the same, for different algorithms the optimal combination of parameters is distinct (see Table \ref{tab_comb} {that shows the number of combinations with the highest accuracies in each algorithm}). 
For instance, {12, 10, and 8} parameter combinations (with the least number of parameters) that achieve the highest accuracy are obtained from ANN, RF, and SVM algorithms, respectively (see Table \ref{tab_resultA} and \ref{tab_comb}). The predictive results are also different. With the predictive model generated by the RF algorithm, {835} BL Lac and {477} FSRQ candidates are diagnosed with a combination of 10 parameters. With the predictive model generated by the ANN algorithm, {868} BL Lac and {444} FSRQ candidates are diagnosed with a combination of seven parameters. With the predictive model generated by the SVM algorithm, {858, 859, 817, 816, 812, 816, 817, 817, 815, 817, 819, 817, 816, 818, or 819} BL Lac style candidates and {454, 453, 495, 496, 500, 496, 495, 495, 497, 495, 493, 495, 496, 494, or 493} FSRQ candidates are diagnosed with a combination of 11 parameters.

We also should note that for the SVM algorithm, there are {15} combinations with {8} parameters that achieve the same maximum accuracy (accuracy={0.918}). 
The diagnosis results are not consistent for different combinations with the same number of parameters and the same accuracy. 
For instance, in the SVM algorithm, the {15} combinations with {8} parameters that achieve the same highest accuracy (accuracy={0.918}), result in different forecasts: {858, 859, 817, 816, 812, 816, 817, 817, 815, 817, 819, 817, 816, 818, or 819} BL Lac and {454, 453, 495, 496, 500, 496, 495, 495, 497, 495, 493, 495, 496, 494, or 493} FSRQ candidates, respectively (see Table \ref{tab_resultA}). When we combine the results of these {454, 453, 495, 496, 500, 496, 495, 495, 497, 495, 493, 495, 496, 494, or 493} combinations, {795 BL Lac and 432} FSRQ candidates are diagnosed; {85} have inconsistent diagnoses. 
{
The first two predictions: {858, or 859} BL Lac and {454, 453} FSRQ candidates (see Table 2), are slightly different from the predictions of the remaining 13 groups.
When we combine the results of the 13 combinations (ignore these two groups), {811 BL Lac and 491} FSRQ candidates are diagnosed; only {11} have inconsistent diagnoses. 
When we combine the results of the RF and ANN algorithms, one of the SVMs (one of these 13 combinations with italic boldface letter, see Table \ref{tab_resultA}) demonstrate {728} BL Lac type candidates and {352} FSRQ type candidates are consistently predicted. However, there are {232} uncertain type BCUs, for which there are inconsistent results.}
Based on the results of this work, a single optimal combination of parameters is not revealed (see Table \ref{tab_comb}). 
Although, {the maximum accuracy of RF algorithms is relatively higher than that of the SVM and ANN algorithms;}
the number of the optimal parameters and combinations in RF and ANN algorithms is relatively less than that of the SVM algorithm.
This implies the interactions among different parameters and/or combinations and different algorithms may be leading to different diagnostic results. 
Therefore, the challenging questions on how to find the optimal parameters and combinations and how to explore the optimal algorithm remain open.

\begin{deluxetable*}{ccccccccch}
\tablenum{4}
\tablecaption{The number of the optimal parameters and combinations\label{tab_comb}}
\tablewidth{0pt}
\tablehead{
\colhead{Classifier} & \colhead{8 par}   & \colhead{9 par}  & \colhead{10 par} &\colhead{11 par}& 
                              \colhead{12 par} & \colhead{15 par} & \colhead{16 par} &\colhead{17 par} &
}
\decimalcolnumbers
\startdata
RF      &     &       & 1     &      &        &        &       &    &    \\
ANN   &     &       &       &      & 1      &        &       &     &    \\
SVM   &15  &36   & 65   &13   & 1     &30     & 8    & 2  &    \\
\enddata
\tablecomments{
The classifiers are presented in Column 1. The number of combinations with the highest accuracies in each algorithm for 7, 9-16 parameters are presented in Columns 2-10. (see the machine-readable format of Table 2).}
\end{deluxetable*}

\begin{table*}
\tablenum{5}
	\centering
	\caption{Median and mean of 724 FSRQ,  332 BL Lacs and 256 unks}
	\label{tab_Median}
	\begin{tabular}{clcccccccccccc} 
		\hline\hline
Paramaters &  Selected                  &\multicolumn{3}{c}{{Median Value}}   & &\multicolumn{3}{c}{Average Value}     \\
\cline{3-5} \cline{7-9}				
{Label}&{Paramaters} & {$\rm M_{fsrq}$} &{$\rm M_{unk}$}& {$\rm M_{bll}$}&~& {$\rm A_{fsrq}$} &$\rm A_{unk}$& {$\rm A_{bll}$} \\
\hline
1	&	PL\_Index        	&	2.57  	&	2.37  	&	2.04  	&	&	2.59  	&	2.36  	&	2.03 	\\
2	&	LP\_Index        	&	2.54  	&	2.33  	&	1.97  	&	&	2.52  	&	2.24  	&	1.92 	\\
3	&	Pivot\_Energy    	&	821.84 	&	1395.26 	&	3054.70 	&	&	840.27 	&	1541.53 	&	3885.54 	\\
4	&	PLEC\_Flux\_Density&	1.00E-12	&	1.87E-13	&	2.88E-14	&	&	2.16E-12	&	3.37E-13	&	5.82E-14	\\
5	&	LP\_Flux\_Density	&	1.04E-12	&	1.91E-13	&	2.89E-14	&	&	2.17E-12	&	3.43E-13	&	5.93E-14	\\
6	&	PL\_Flux\_Density	&	8.62E-13	&	1.63E-13	&	2.60E-14	&	&	1.88E-12	&	3.02E-13	&	5.33E-14	\\
7	&	Frac2\_Variability	&	0.52  	&	0.00  	&	0.00  	&	&	0.53  	&	0.23  	&	0.05 	\\
8	&	nuFnu\_Band7  	&	3.97E-17	&	1.38E-16	&	1.99E-13	&	&	2.30E-14	&	7.34E-14	&	3.23E-13	\\
9	&	PLEC\_Index    	&	2.11   	&	2.03  	&	1.75  	&	&	2.01  	&	1.71   	&	1.65 	\\
10	&	Flux\_Band2     	&	6.88E-09	&	2.68E-09	&	1.20E-09	&	&	9.01E-09	&	4.25E-09	&	2.07E-09	\\
11	&	Flux\_Band7     	&	9.65E-16	&	4.27E-15	&	3.90E-12	&	&	5.85E-13	&	1.65E-12	&	6.00E-12	\\
12	&	nuFnu\_Band2  	&	1.60E-12	&	6.35E-13	&	2.91E-13	&	&	2.10E-12	&	9.93E-13	&	4.92E-13	\\
13	&	Frac\_Variability	&	0.51  	&	0.30  	&	0.18  	&	&	0.54  	&	0.37  	&	0.24 	\\
14	&	PLEC\_Expfactor	&	5.00E-03	&	3.02E-03	&	9.80E-04	&	&	7.79E-03	&	5.04E-03	&	1.58E-03	\\
15	&	Variability2\_Index	&	75.80 	&	49.38 	&	43.74 	&	&	138.09 	&	105.96 	&	47.45 	\\
16	&	Variability\_Index	&	20.62 	&	10.15 	&	9.04  	&	&	56.79 	&	45.15 	&	11.02 	\\
17	&	nuFnu\_Band6   	&	4.41E-14	&	1.09E-13	&	3.71E-13	&	&	1.30E-13	&	2.17E-13	&	4.68E-13	\\
18	&	Flux\_Band3     	&	1.76E-09	&	9.01E-10	&	4.86E-10	&	&	2.34E-09	&	1.40E-09	&	6.88E-10	\\
19	&	Npred              	&	690.34 	&	402.20 	&	231.29 	&	&	832.93 	&	495.96 	&	289.25 	\\
20	&	Flux\_Band6     	&	1.96E-12	&	4.68E-12	&	1.55E-11	&	&	5.81E-12	&	9.47E-12	&	1.95E-11	\\
21	&	nuFnu\_Band3  	&	1.13E-12	&	5.84E-13	&	3.31E-13	&	&	1.53E-12	&	9.34E-13	&	4.66E-13	\\
22	&	Flux\_Band1     	&	7.03E-09	&	1.91E-10	&	1.98E-10	&	&	1.23E-08	&	6.79E-09	&	5.17E-09	\\
23	&	nuFnu\_Band1   	&	1.12E-12	&	3.02E-14	&	3.17E-14	&	&	1.95E-12	&	1.09E-12	&	8.28E-13	\\
\hline
	\end{tabular}
	\\
\tablecomments{Column 1 presents the parameter labels in the sample. Column 2 lists the selected parameters.
The  median value   of FSRQs ($\rm M_{fsrq}$), uncertain type BCUs ($\rm M_{unk}$) and  BL Lacs ($\rm M_{bll}$) are presented in Columns 3, 4,  and 5, respectively. 
The  average value    of FSRQs ($\rm A_{fsrq}$), uncertain type BCUs ($\rm A_{unk}$) and  BL Lacs ($\rm A_{bll}$) are presented in Columns 6, 7,  and 8, respectively. 
}
\end{table*}

\begin{figure*}[bpt]
\centering
 \includegraphics[height=6cm,width=8cm]{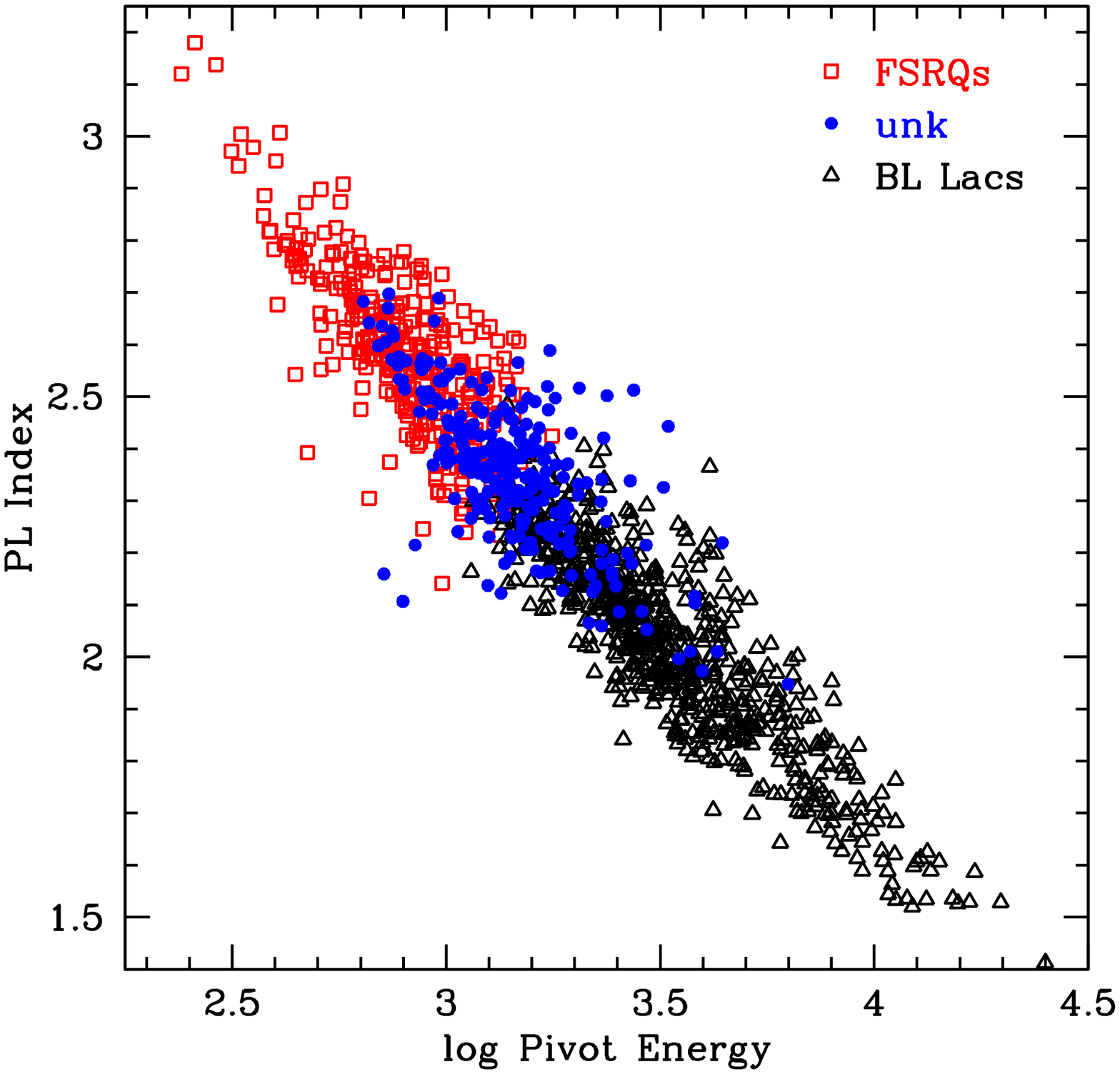}
 \includegraphics[height=6cm,width=8cm]{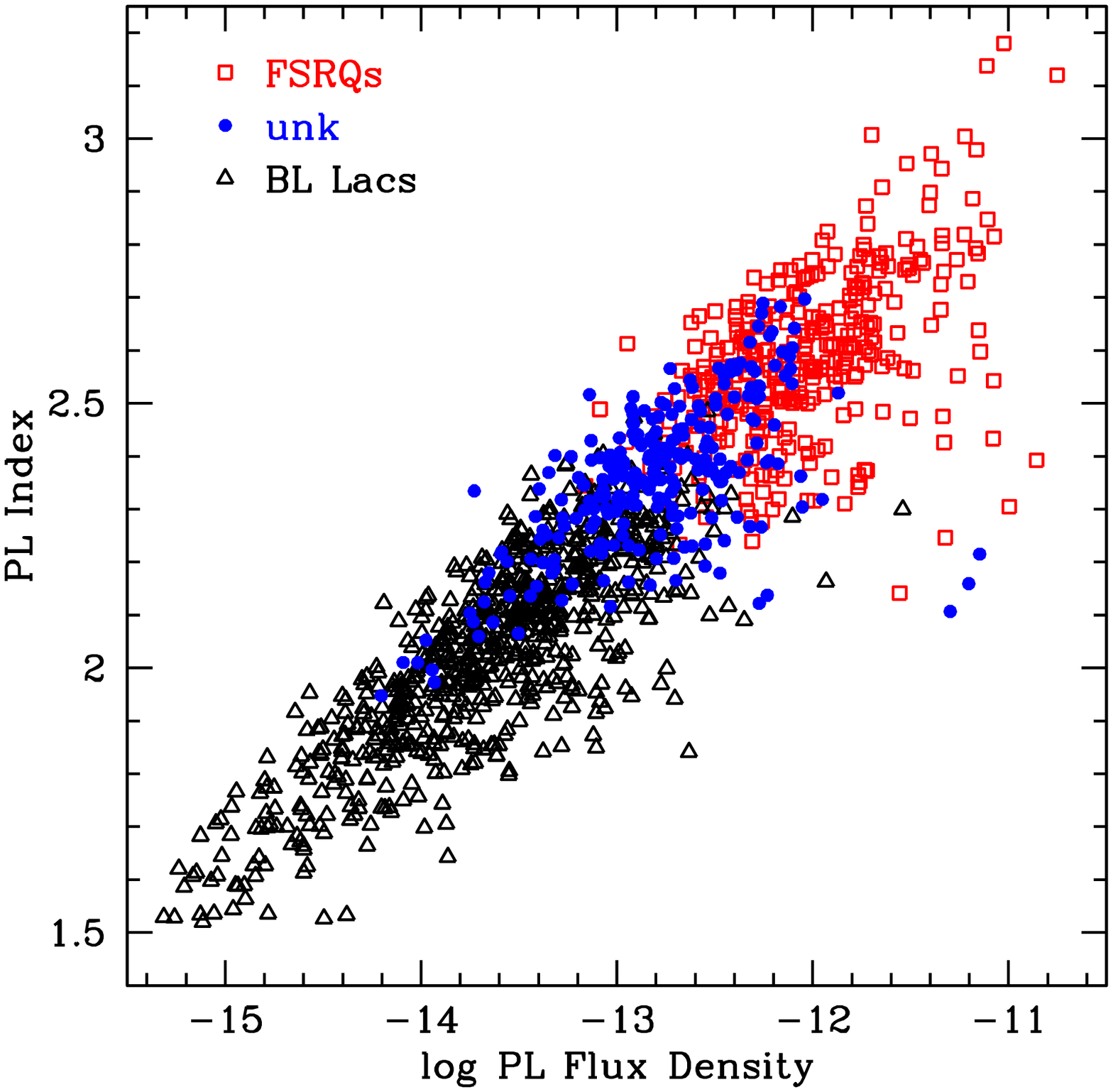}
 \includegraphics[height=6cm,width=8cm]{figa03_empty.eps} 
 \includegraphics[height=6cm,width=8cm]{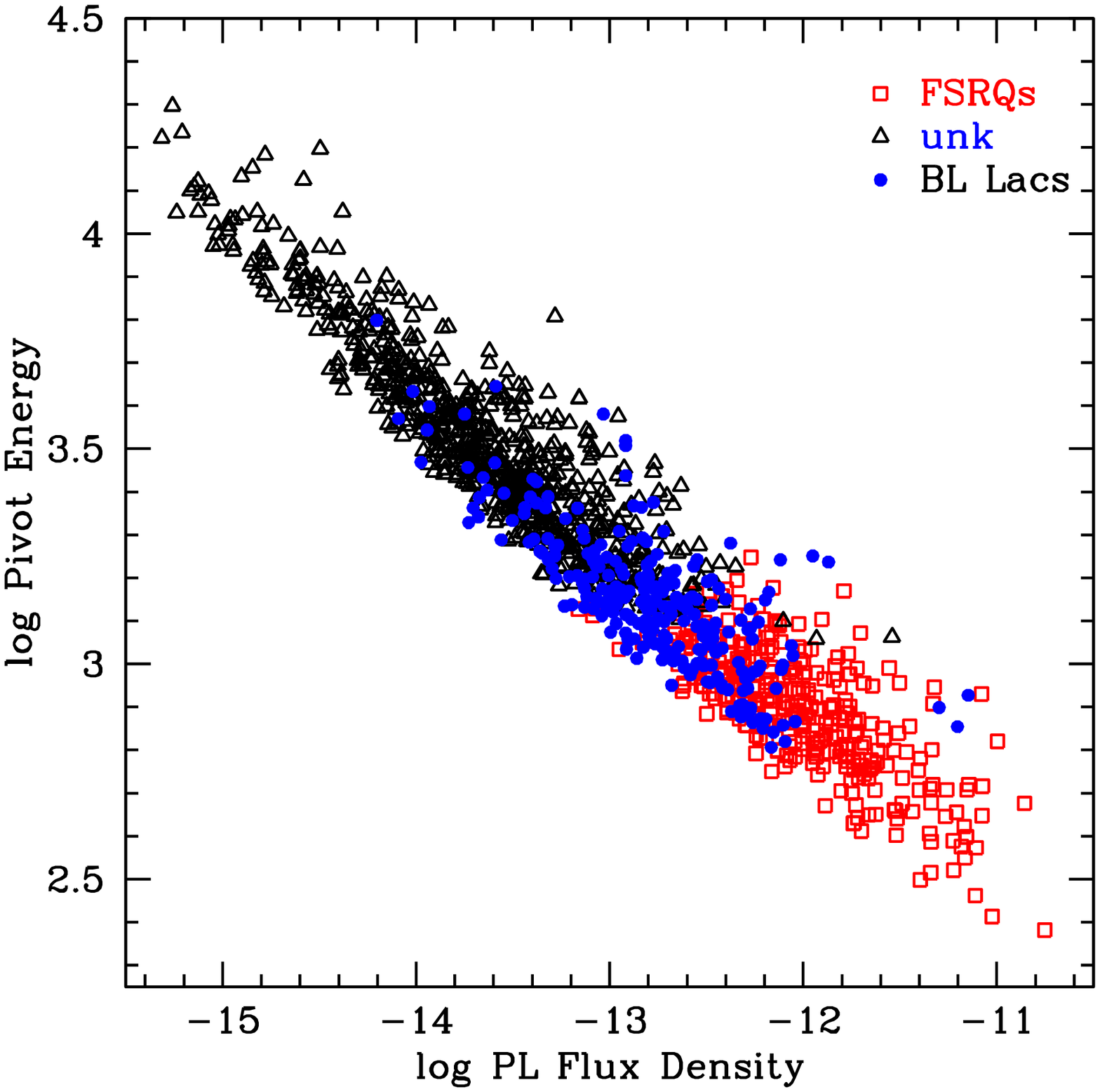}
\caption{(Color online) Classification scatter diagram for a subset of the parameters (PL\_Index, Pivot\_Energy and PL\_Flux\_Density).
Only a  portion is shown here for guidance regarding its form and content. 
The red empty squares,  black empty triangles,   and blue solid circles indicate FSRQs, BL Lacs and uncertain type BCUs respectively. 
}
 \label{sub_figb}
\end{figure*}

{When comparing the 256 uncertain type BCUs with the 724 BL Las and the 332 FSRQs predicted in the work, 
we find that the FSRQs have a larger median value and average value for most parameters (e.g., ``PL\_Index",  ``LP\_Flux\_Density", or ``Flux\_Band2", etc. see Table \ref{tab_Median}) compared with BL Las.
The remaining parameters (``Pivot\_Energy", ``nuFnu\_Band7", ``Flux\_Band7", ``nuFnu\_Band6", ``Flux\_Band6") show a smaller median value and average value for the FSRQs.
The median value and average value of the uncertain type BCUs can be considered in two cases. 
The first case has smaller values than those of the FSRQ and greater than those of the BL Las
({$\rm M_{fsrq} >  M_{unk}> \rm M_{bll}$},  and {$\rm A_{fsrq}  >   A_{unk}  >  A_{bll}$}).
The second case has values greater than those of the FSRQ and smaller than those of the BL Las
({$\rm M_{fsrq} <  M_{unk} < \rm M_{bll}$},  and {$\rm A_{fsrq}  <  A_{unk}  <  A_{bll}$}, see Table \ref{tab_Median}).
The median value and average value of the uncertain type BCUs always locate between that of FSRQ and BL Las.
In addition, from the scatter plots (e.g., see Figure \ref{sub_figb}), we can easily find that these unk sources are usually located in overlapping regions of the FSRQ and BLLac distributions. 
This may make these unk sources difficult to distinguish and make them uncertain.}

%\clearpage

Cross-matching  the 4FGL predictions ({1312} BCUs) obtained from  the combined the results of the three algorithms used in the work 
and the 3FGL predictions (400 BCUs), we obtained 219 common objects (see Table \ref{tab_comp}).
Among the common objects, 119 BL Lacs, 66 FSRQs candidates, and 34 unks (still no explicit predictions) were predicted in this work (4FGL prediction).
In addition, 113 BL Lacs, 47 FSRQs candidates, and 59 unks were previously predicted in the 3FGL work for the results of combining four algorithms (paper I).
For the 113 BL Lacs candidates predicted in the 3FGL work (paper I),  most of which (97 sources, approximately 85.84\%, are consistent with the results of the 3FGL) were also predicted as BL Lacs candidates, only three sources are predicted as FSRQs candidates; 13 sources are remain without a clear prediction in the work.
For the 47 FSRQs candidates reported in the 3FGL work (paper I), most of which (40 sources, approximately 85.84\%, are consistent with the results of 3FGL) were also predicted as FSRQs candidates, non-source is predicted as BL Lac candidates, and seven sources are remain without a clear prediction in the work.
Overall, approximately 85\% of the predicted results are consistent with the results of the 3FGL, which shows that the 3FGL and 4FGL classifications are basically consistent.
Also, for the 60 unks predicted in the 3FGL work (paper I), 
there are 22 and 23 sources are classified as  BL Lac and FSRQ candidates, respectively; 
there are still 14 sources are not provided with a clear prediction in the work.
In addition, we also compared the results of each algorithm of the previous  work (Paper I); these results are reported in Table \ref{tab_comp}.

Cross-matching the 3FGL predictions, the BL Lacs and FSRQs have been spectrally verified in the 4FGL ({694  FSRQs and 1131 BL Lacs}),
140 sources are  verified as BL Lacs and 19 sources are  verified as FSRQs in the 4FGL catalog (see Table \ref{tab_comp}); 
125 BL Lacs, 13 FSRQs candidates, and 21 unks were predicted in the 3FGL work for the results of combining four algorithms (paper I).
The 140 BL Lacs have been spectrally verified in the 4FGL,  most of which (120 sources, approximately 85.71\% (120/140), are consistent with the 3FGL predictions) were also predicted as BL Lacs candidates, only 3 sources are predicted as FSRQs candidates; 17 sources are without a clear prediction in 3FGL work.
For the 19 FSRQs that have been spectrally verified in the 4FGL, approximately 1/2 sources (10 sources, are consistent with the 3FGL predictions) were also predicted as FSRQs candidates, 5 sources are predicted as BL Lac candidates; 4 sources remain without a clear prediction in the 3FGL work, respectively.
The prediction accuracy of BL Lacs is relatively high, while the prediction accuracy of FSRQs is lower.
For the results of each algorithm of the 3FGL work (Paper I), similar results are also reported (see Table \ref{tab_comp}).
However, the predictions obtained by applying various methods exhibit a higher prediction accuracy (e.g., 120/125, approximately 96\% for BL Lacs' prediction; 10/13,  approximately 76.9\% for FSRQs' prediction).
These results show that our classification prediction approach is very reliable.

\begin{deluxetable*}{lcc|ccccccccch}
\tablenum{6}
\tablecaption{Comparison of 3FGL predictions and 4FGL results\label{tab_comp}}
\tablewidth{0pt}
\tablehead{
&&\multicolumn{1}{c}{{3FGL predictions}}   &  & \multicolumn{3}{c}{{4FGL predictions}} 
                                                         &  & \colhead{3FGL predictions}&&\multicolumn{2}{c}{{4FGL verified}}  \\
\cline{3-3} \cline{5-7}	                     \cline{9-9}  \cline{11-12}
\colhead{algorithm} &\colhead{Class} &\colhead{N$^a$} && \colhead{bll}   & \colhead{fsrq}  & \colhead{unk}   
        &&\colhead{N$^b$} & &\colhead{bll}   & \colhead{fsrq}  &                     
}
%\decimalcolnumbers
\startdata
               &N$^c$  &  219      && 119     & 66      & 34        &&159      &&140  & 19   &     \\
\cline{2-13} 
               &bll      &  113        &&  97     &  3       &  13       &&125      &&120   & 5    &      \\
4 methods&fsrq    &  47         &&   0     &  40      &  7        &&13        &&3     & 10    &     \\
               &unk    &   59        &&  22     &  23     &  14       &&21        &&17    & 4    &     \\
\hline
Mclust     &bll      &  145        &&  108     &  17      &  20      &&137     &&129   &8     &      \\
               &fsrq    &  74        &&   11      &  49      &  14       &&22      &&11     & 11     &       \\
\hline
rpart        &bll      &  137        &&  103     &  12      &  22      &&131    &&125    & 6      &      \\
               &fsrq    &  82        &&   16      &  54      &  12       &&28    &&15      & 13      &       \\
\hline
RF           &bll      &  149        &&  118     &  9      &  22      &&143    &&135    &8     &      \\
               &fsrq    &  70        &&   1       &  57      &  12       &&16      &&5       &11     &       \\
\hline
SVM        &bll      &  148        &&  117     &  10      &  12      &&141     &&135      & 6     &      \\
               &fsrq    &  71        &&   2       &  56      &  6       && 18      &&5         & 13      &       \\
\hline
\enddata
\tablecomments{The classifiers and classes are presented in Column 1 and 2. 
Column 3-6 shows the results of comparison of the 3FGL predictions and the 4FGL predictions for common objects.
The results of comparison of the 3FGL predictions, BL Lacs and FSRQs verified in the 4FGL are presented in Columns 7-9.
Where, $^a$ and $^b$ indicate the number of the 3FGL predictions; $^c$ indicates the number of 4FGL sources in the cross-matching the 3FGL predictions and 4FGL predictions,  or the sources verified in the 4FGL.}
\end{deluxetable*}

In addition, we also should note that all of the default settings {(e.g., the probabilities: p $>$ 0.5 in each classifier to consider a correct classification, see Table \ref{ML_result})} for each of the three classification functions (e.g., randomF orest(), svm() and nnet() function) are used in Section 4. For each different classification method, choosing the calculation model and setting each parameter in the fitting function can also affect the predictive models, accuracy, and results (e.g., see the  discussions in \citealt{2019ApJ...872..189K}). The selections of the appropriate parameter settings need further investigation; this is  beyond the scope of the current work. 
{
Also, we should note that approximately 4/5 and 1/5 of the known classification blazars are assigned as the training data set and validation data set, respectively. The choice of 4/5 and 1/5  is a bit arbitrary. The predictive accuracy and results may be affected by the training data set and validation data set. This issue has been discussed in Paper I (see page 8 in Kang et al. 2019 for the details and discussions).}
Similar or identical parameters (see Table \ref{tab:test}), such as PL\_Index, LP\_Index and PLEC\_Index; 
or PL\_Flux\_Density,  LP\_Flux\_Density and PLEC\_Flux\_Density;
or Variability\_Index and Variability2\_Index,
are also used in the SML at the same time, which should be cautioned.
In addition, it must be highlighted that, in this work, the sample selection methods used (\citealt{2018RAA....18...56K,2019ApJ...872..189K}),  may affect the source distributions and the results of the analysis. However, this work provides some clues in applying SML algorithms to evaluate the classification of Fermi BCUs.

Finally, it should be noted that in this work, only a subset of the parameters (23 out of 33 parameters) have been selected to search for the optimal parameter combination based on the preliminary version of the data (see \citealt{2019arXiv190210045T}), which may cause bias. More variables and the final complete sample are needed to further test and address the issue.

%\acknowledgments
\section*{Acknowledgements}
We thank the anonymous referee for very constructive and helpful comments and suggestions, which greatly helped us to improve our paper.
This work is  partially supported by the National Natural Science Foundation of China (Grant Nos.11763005, 11873043, U1931203, 11733001, 11622324, U1531245, and 11573009),
the Research Foundation for Scientific Elitists of the Department of Education of Guizhou Province (QJHKYZ[2018]068), 
the Science and Technology Foundation of Guizhou Province (QKHJC[2019]1290), 
the Science and Technology Platform and Talent Team Project of Science and Technology  Department of Guizhou Province (QKH $\ast$ Platform \& Talent[2018]5777, [2017]5721), 
and the Research Foundation of Liupanshui Normal University (LPSSY201401, LPSSYSSDPY201704, LPSZDZY2018-03, LPSSYZDXK201801, and LPSSYsyjxsfzx201801).


%\bibliography{Optimal_4FGL_23_sm}{}
%\bibliographystyle{aasjournal}

%% \allauthors

%% \listofchanges

 \end{CJK*}
\end{document}